\documentclass[%
 reprint,
superscriptaddress,
 amsmath,amssymb,
 aps,
prl,
]{revtex4-2}
\usepackage{graphicx}% Include figure files
\usepackage{dcolumn}% Align table columns on decimal point
\usepackage{bm}% bold math
\usepackage[usenames,dvipsnames]{xcolor}
\usepackage{lipsum}
\usepackage{physics}
\usepackage{multirow}
\usepackage{float}

\newcommand{\pdagger}{\phantom{\dagger}}

\definecolor{hred}{HTML}{E43E4C}
\usepackage{hyperref}
\hypersetup{
    pdfstartview={FitH},    % fits the width of the page to the window
    colorlinks=true,       % false: boxed links; true: colored links
    linkcolor=blue,          % color of internal links (change box color with linkbordercolor)
    citecolor=blue,        % color of links to bibliography
    filecolor=magenta,      % color of file links
    urlcolor=blue           % color of external links
}

\usepackage{silence}
\begin{document}

\preprint{APS/123-QED}

\title{Supersolidity and Simplex Phases in Spin-1 Rydberg Atom Arrays}

\newcommand{\HarvardPhysicsAddress}{Department of Physics, Harvard University, Cambridge, Massachusetts 02138, USA}
\newcommand{\PrincetonAAddress}{Department of Physics, Princeton University, Princeton, NJ 08544, USA}
\newcommand{\PrincetonBAddress}{Princeton Center for Theoretical Science, Princeton University, Princeton, NJ 08544, USA}

\author{Vincent S. Liu}
\affiliation{\HarvardPhysicsAddress}

\author{Marcus Bintz}
\affiliation{\HarvardPhysicsAddress}

\author{Maxwell Block}
\affiliation{\HarvardPhysicsAddress}

\author{Rhine Samajdar}
\affiliation{\PrincetonAAddress}
\affiliation{\PrincetonBAddress}

\author{Jack Kemp}
\affiliation{\HarvardPhysicsAddress}

\author{Norman Y. Yao}
\affiliation{\HarvardPhysicsAddress}

\date{\today}% It is always \today, today,

%%%%%%%%%%%%%%%%%%%%%%%%%%%%%%%%%%%%%
% Abstract
%%%%%%%%%%%%%%%%%%%%%%%%%%%%%%%%%%%%%

\begin{abstract}
% keywords:
% - quantum phases of matter
% - density matrix renormalization group
% - rydberg atom arrays
% - spin-1
% - supersolid
% - simplex phase
% - two-dimensional

Neutral atoms become strongly interacting when their electrons are excited to loosely bound Rydberg states.
We investigate the strongly correlated quantum phases of matter that emerge in two-dimensional atom arrays where three Rydberg levels are used to encode an effective spin-1 degree of freedom. 
Dipolar exchange between such spin-1 Rydberg atoms naturally yields two distinct models: (i) a two-species hardcore boson model, and (ii) upon tuning near a F\"orster resonance, a dipolar spin-1 XY model. 
Through extensive, large-scale infinite density matrix renormalization group calculations, we provide a broad roadmap predicting the quantum phases that emerge from these models on a variety of lattice geometries: square, triangular, kagome, and ruby.
We identify a wealth of correlated states, including lattice supersolids and simplex phases, all of which can be naturally realized in near-term experiments.

\end{abstract}


%%%%%%%%%%%%%%%%%%%%%%%%%%%%%%%%%%%%%
% Make header & TOC
%%%%%%%%%%%%%%%%%%%%%%%%%%%%%%%%%%%%%

\maketitle

%%%%%%%%%%%%%%%%%%%%%%%%
% INTRODUCTION
%%%%%%%%%%%%%%%%%%%%%%%%

Strongly interacting qudit ensembles can exhibit a rich variety of many-body phenomena.
The complexity enabled by a large local Hilbert space pays dividends for error correction~\cite{ketkarNonbinaryStabilizerCodes2006, campbellMagicStateDistillationAll2012, yoshidaInformationStorageCapacity2013}, quantum metrology~\cite{ciampiniQuantumenhancedMultiparameterEstimation2016,shlyakhovQuantumMetrologyTransmon2018}, and the simulation of fast-scrambling dynamics~\cite{Gu:2017,Nahum:2017,zhuangScramblingComplexityPhase2019,blokQuantumInformationScrambling2021}.
In the context of phases of matter, higher-spin models are  crucial for the simulation of magnetic materials~\cite{Renard:1987,gao2020spin, Chamorro2018, Chauhan:2020, Quilliam2016, Zhang2019} and can also naturally guide one to more exotic physics.
Most famously, promoting the antiferromagnetic Heisenberg spin-1/2 chain to spin-1 opens a gap which protects a Haldane phase with hidden string order~\cite{haldane:1983,affleck:1987}. 
Such higher-spin chains have been experimentally realized in a number of quantum simulation platforms~\cite{senkoRealizationQuantumIntegerSpin2015,Chung:2021, mogerleSpinHaldanePhase2022} and remain the subject of intense theoretical interest~\cite{Chen2003, Lauchli:2006, Pollman2012, kjallPhaseDiagramAnisotropic2013,gilsAnyonicQuantumSpin2013, Gong2016}. 

The combination of higher spin and higher dimensions can favor even more complex low-temperature orders.
For example, it permits novel forms of continuous symmetry breaking, such as nematic phases that break spin-rotation symmetry while preserving time-reversal symmetry~\cite{papanicolaou:1988,harada:2002, Lauchli:2006, manmanaPhaseDiagramContinuous2011}.
Moreover, a large local Hilbert space can favor the emergence of long-range entangled quantum spin liquid ground states, including those hosting non-Abelian anyons~\cite{bieri:2012, xu:2012, serbyn:2011, buessen:2018, PhysRevB.105.155104}.
Despite these intriguing possibilities, the development of programmable quantum simulators capable of studying higher-spin models in $D$\,$>$\,$1$ remains nascent~\cite{gorshkovQuantumMagnetismPolar2011,Chung:2021, ringbauerUniversalQuditQuantum2022, Fu:2022, Roy:2023, weitenberg:2011,cappelliniDirectObservationCoherent2014,paganoOnedimensionalLiquidFermions2014,scazza:2014,zhang:2014,miranda2015site,yamamoto2016ytterbium,miranda2017site,sonderhouse:2020,schine2021long}.
One route to such investigations is provided by Rydberg atom arrays, which have achieved striking success in the realization of spin-1/2 phases in arbitrary 2D geometries~\cite{yao:2013, glaetzle:2014,  samajdar:2020, celi:2020, ebadiQuantumPhasesMatter2021, semeghini:2021, giudici:2022, chenContinuousSymmetryBreaking2023}, as well as the coherent control of multiple, distinct Rydberg states~\cite{lienhardRealizationDensityDependentPeierls2020,chenStronglyInteractingRydberg2024}. 
These capabilities can be naturally synthesized to realize higher-spin Rydberg models in 2D, a subject of ongoing experimental effort~\cite{lienhardRealizationDensityDependentPeierls2020, Kanungo:2022, chenStronglyInteractingRydberg2024}.
Predicting the correlated many-body states that might emerge in such systems remains an essential open question~\cite{Weber:2022, Ohler:2022, zhangSimulatingTwocomponentBoseHubbard2024, homeierAntiferromagneticBosonicModels2024}. 

%%%%%%%%%%%%%%%%%%%%%%%%%
% FIGURE 1
%%%%%%%%%%%%%%%%%%%%%%%%%
\begin{figure}%[H]
    \centering
    \includegraphics[width=1.0\columnwidth]{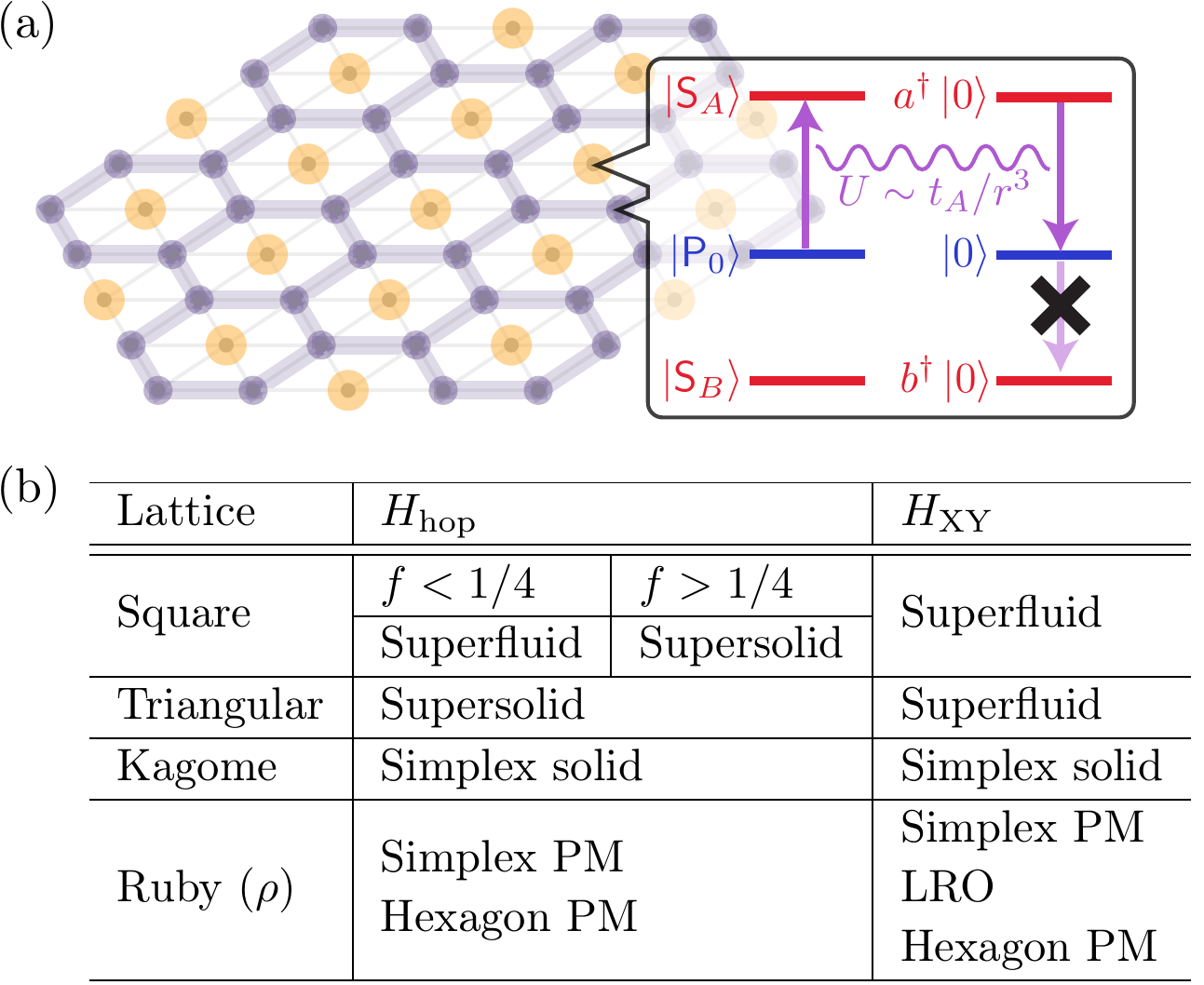}
    \caption{(a) Simple three-level Rydberg encoding scheme using two different $\mathsf{S}$ levels and one $\mathsf{P}$ level under the dipolar exchange interaction. If the $\mathsf{S}$ levels are symmetrically detuned and the transition dipole moments are equal, the model realizes the spin-1 XY Hamiltonian. (b) Table summarizing various phases of $H_{\rm AB}$ and $H_{\rm XY}$ on relevant lattices. $f \equiv f_A = f_B$ refers to the total filling fraction of bosons. The ruby lattice is investigated for multiple aspect ratios $\rho$, with the phases listed in order of decreasing aspect ratio. Here, ``PM'' and ``LRO'' stand for paramagnet and long-range order, respectively.}
    \label{fig:scheme}
\end{figure}

\begin{figure}%[H]
    \centering
\includegraphics[width=\columnwidth]{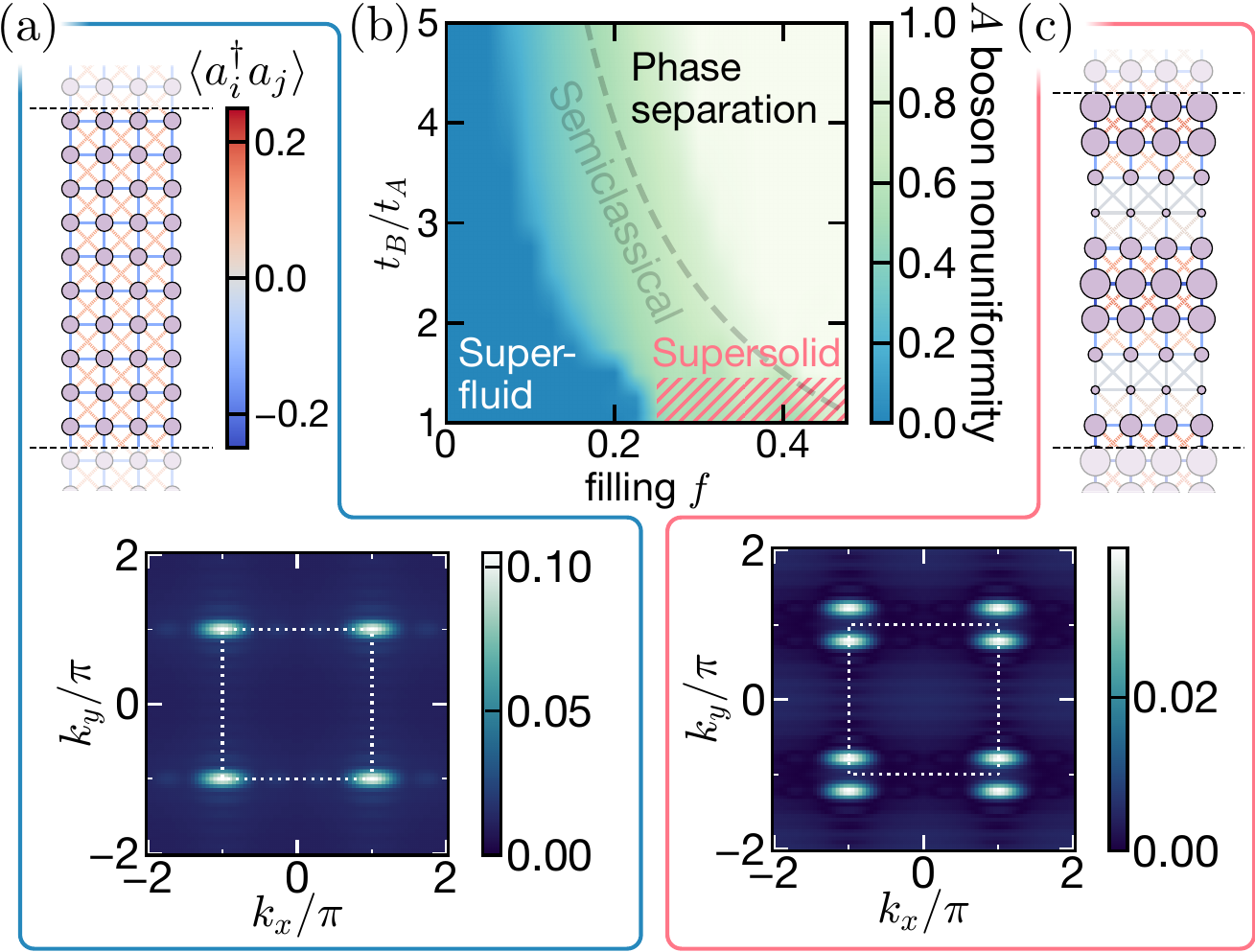}
    \caption{Phases of $H_{\rm AB}$ on the square lattice, showing (a) superfluid and (c) supersolid order, with data from cylinder iDMRG. The phase diagram as a function of the filling $f$ and the hopping ratio $t_B/t_A$ is shown in panel (b), with approximate superfluid, supersolid, and phase-separated regions labeled. It is unknown whether the supersolid order is stable to $t_B/t_A \neq 1$. The dashed line shows the boundary expected from a semiclassical energy argument. Color scale: phase separation nonuniformity order parameter $\Delta N_A = \mathrm{max} \left( N_A \right) - \mathrm{min} \left( N_A \right)$, with $N_A$, $N_B$ denoting the onsite boson densities of $A$ and $B$ bosons ($a^\dagger a$, $b^\dagger b$ respectively). (a,c) Top: real-space plots of the superfluid and supersolid phases respectively, with the circle size representing onsite densities of $A$ bosons and bonds representing off-diagonal $\langle a_i^\dagger a_j \rangle$ correlations. Bottom: structure factor $| \sum_{ij} \langle a_i^\dagger a_j e^{i \mathbf{k} \cdot \mathbf{r}_{ij}} \rangle |$ for the superfluid and supersolid phases. While the superfluid phase has the structure factor peaked at the corners of the Brillouin zone, the supersolid phase has doubled and slightly shifted peaks in the $y$ (periodic) direction at $\left( \pm \pi, \pm \pi \pm \Delta k \right)$. In the supersolid phase, the $A$ and $B$ boson density is nonuniform (with oscillations $\sim \sin \left( 2 \Delta k y \right)$) but the hole density is uniform; for phase separation, $A$ bosons separate from $B$ bosons and holes, which remain uniform with respect to each other~\cite{supp}.
    }
    \label{fig:phasediagram}
\end{figure}

In this Letter, we consider the pair of spin-1 models that most naturally emerge from the dipolar interactions of a three-state Rydberg encoding; the first is equivalent to a simple model of two species of hardcore bosons with long-range hopping (Fig.~\ref{fig:scheme}), while the second corresponds to the spin-1 dipolar XY Hamiltonian.
We perform extensive, large-scale, infinite density matrix renormalization group (iDMRG) computations~\cite{tenpy} to chart the ground-state phase diagrams (Fig.~\ref{fig:phasediagram}) of both models on a diverse set of lattice geometries [Fig.~\ref{fig:scheme}(b)]. 
Our main predictions are twofold.
First, we find that while the spin-1 dipolar XY model generically exhibits superfluidity on the square and triangular lattices, the dual-species boson model may instead host lattice supersolids, where spin-density-wave order accompanies the superfluidity.
Second, we show that the same models on the kagome and ruby lattices instead stabilize simplex phases in which spins strongly trimerize about lattice triangles---an intrinsically spin-1 phenomenon that generalizes the formation of valence bonds in spin-1/2 systems.
Finally, we demonstrate that all of the above phases can be realized in parameter regimes readily accessible to current-generation experiments.

%%%%%%%%%%%%%%%%%%%%%%%%%%%%%%%%
% MODEL HAMILTONIANS
%%%%%%%%%%%%%%%%%%%%%%%%%%%%%%%%
\emph{Spin-1 Rydberg Hamiltonians.}---Consider an array of neutral atoms with an effective spin encoded in three distinct Rydberg levels: for example, two $\mathsf{S}$ states ($\ket{\mathsf{S}_A}$ and $\ket{\mathsf{S}_B}$) coupled to a single $\mathsf{P}$ state $\ket{\mathsf{P}_0}$, as shown in Fig.~\ref{fig:scheme}(a)~\cite{homeierAntiferromagneticBosonicModels2024}.
In the Rydberg manifold, the large transition dipole moments, $\mu_{A,B}$, between the $\mathsf{S}_{A,B}$ and $\mathsf{P}$ states lead to strong dipolar exchange interactions.
Depending on the energy differences between the Rydberg states, $\delta_A = |E_A-E_0|$, $\delta_B = |E_B-E_0|$, there are two distinct effective Hamiltonians that may govern the many-body physics. 

Generically, $\delta_A \ne \delta_B$, and the dipolar  interactions are  significantly weaker than the energy difference $\delta_A - \delta_B$.
Energy conservation (i.e., taking the secular approximation) then enforces that the number of atoms in the $\ket{\mathsf{S}_A}$ and $\ket{\mathsf{S}_B}$ states are each conserved.
In this case, the relevant two-body interactions are resonant exchange processes taking the form,
    $H_{\rm int}= | \mu^{}_{A} |^2  \ket{\mathsf{S}_A \mathsf{P}_0}\bra{\mathsf{P}_0 \mathsf{S}_A} + | \mu^{}_{B} |^2 \ket{\mathsf{S}_B \mathsf{P}_0}\bra{\mathsf{P}_0 \mathsf{S}_B} + \mathrm{h.c.}$\,.
It is now natural to reformulate this Hamiltonian in terms of two species of hardcore bosons, labeled $A$ and $B$, hopping on the lattice.
Denoting the creation operators for these bosons as $a^\dagger$ and $b^\dagger$, respectively, the three Rydberg levels can be identified with the three possible occupations of a single site: $\ket{\mathsf{P}_0} \equiv \ket{0}$, $\ket{\mathsf{S}_A} \equiv a^{\dagger} \ket{0}$, and $\ket{\mathsf{S}_B} \equiv b^{\dagger} \ket{0}$.
In this language, the many-body Hamiltonian reduces to that of hopping hardcore bosons,
\begin{align}
\label{eq:Hhop}
    H^{}_{\rm AB} = \sum_{i,j } \left[ \frac{t_A}{r_{ij}^3}  a_i^\dagger a^{\pdagger}_j  + \frac{t_B}{r_{ij}^3}  b_i^\dagger b^{\pdagger}_j + \mathrm{h.c.}  \right],
\end{align}
where $|t_{A,B}|$\,$\sim$\,$\left| \mu_{A,B} \right|^2$ and $r_{ij}$ denotes the distance between sites $i$ and $j$.
While both signs of the hopping can be experimentally realized~\cite{geierTimereversalDipolarQuantum2024}, here, we focus on the antiferromagnetic case, $t_{A,B}>0$, which leads to a richer variety of phases.
Owing to the aforementioned energy conservation,
$H_\textrm{AB}$ exhibits a continuous ${{\rm U}(1)} \times {{\rm U}(1)}$ symmetry with two conserved charges corresponding to the filling fractions, $f_A$ and $f_B$, of the $A$ and $B$ bosons. 
%; it also respects any lattice space-group symmetries.

An alternate effective Hamiltonian
arises when there is a F\"orster resonance: $E_A-E_0 \approx E_0 - E_B$.
Then, there is an additional allowed pair-creation process, 
$\mu^{}_{A} \mu^{}_{B}(\ket{\mathsf{S}^{}_B \mathsf{S}^{}_A}\bra{0 0} + \ket{\mathsf{S}^{}_A \mathsf{S}^{}_B}\bra{0 0}) +\mathrm{h.c.}
$
~This  reduces the symmetry from $\mathrm{U}(1)$\,$\times$\,$\mathrm{U}(1)$ to just $\mathrm{U}(1)$; in the hardcore boson language, only the density \textit{difference} $f_A-f_B$, is conserved. 
We note that both the F\"orster resonance and the dipole moments can be tuned by dressing the $\ket{\mathsf{S}_A}$ and $\ket{\mathsf{S}_B}$ levels with different Rydberg states~\cite{vanbijnenQuantumMagnetismTopological2015}. 
This allows us to further specialize to the scenario in which $\mu_{A}= \mu_{B}$, so that the effective Hamiltonian corresponds to the spin-1 dipolar XY model,
\begin{align}
    H^{}_{\rm{XY}} = J \sum_{i, j} \frac{ S_i^x S_j^x + S_i^y S_j^y }{r_{ij}^3},
\end{align}
where $\ket{\mathsf{P}_0}$ encodes the $\ket{m_s = 0}$ state and $\ket{\mathsf{S}_{A,B}}$ the $\ket{m_s = \pm 1}$ states. 
Returning to the language of spins, the conserved $\mathrm{U}(1)$ charge is  the total magnetization, $S^z_\textrm{tot}$, along the $z$-axis.

%%%%%%%%%%%%%%%%%%
% FIGURE 2
%%%%%%%%%%%%%%%%%%

%%%%%%%%%%%%%%%%%%%%%%%%%%
% PHASES (SQUARE+TRIANGLE)
%%%%%%%%%%%%%%%%%%%%%%%%%%
\emph{Superfluid and supersolid phases.}---We now explore the ground-state phase diagrams of $H_{\textrm{AB}}$ and $H_{\textrm{XY}}$, beginning with the square lattice.  
For $H_{\rm AB}$, we will tune the filling fraction, $f\equiv f_A$\,$=$\,$f_B$, and the ratio of the hopping strengths, $t_B/t_A$ [Fig.~\ref{fig:phasediagram}(b)].
Starting with $t_B/t_A = 1$, where the Hamiltonian  exhibits an additional $A$-$B$ boson exchange symmetry, we utilize  iDMRG to compute the ground state on an infinitely long cylinder with circumference $L_y = 10$ sites (see Supplemental Material for  discussions of other circumferences, as well as methodological details~\cite{supp}). 
At low filling fractions,  $f<1/4$, we find that the system exhibits superfluid order  with a uniform real-space density profile and a superfluid order parameter $| \langle a^{\dagger}_i a_j e^{i \mathbf{k} \cdot \mathbf{r}_{ij}} \rangle |$ peaked at the corners of the Brillouin zone indicating off-diagonal long-range order [Fig.~\ref{fig:phasediagram}(a)]. 

For higher filling fractions, $f > 1/4$, we instead observe lattice supersolidity, characterized by a nonuniform density of $A$ and $B$ bosons in real space, but with the hole density remaining uniform.
%[Fig.~\ref{fig:phasediagram}(b)]
In this case, the $A$ and $B$ bosons form separate stripes of constant filling [Fig.~\ref{fig:phasediagram}(c)].
The superfluid order parameter becomes doubly peaked near each corner of the Brillouin zone, with a greater peak separation in Fourier space corresponding to narrower stripes in real space.
As the filling fraction increases, the peaks separate further and the stripes become narrower.
This supersolid phase spontaneously breaks three types of symmetry: $\mathrm{U}(1)$ spin rotation (for both types of bosons), spatial translation and rotation, and $A$-$B$ exchange. It does not, however, break the translation symmetry of the \textit{holes}.
An alternative phase that does spontaneously break this hole-translation symmetry is found upon explicitly breaking the $A$-$B$ exchange symmetry. 
In particular, if $t_B \gg t_A$, then the lowest energy configuration is one in which the $A$ bosons precipitate out, allowing the remaining $B$ bosons and holes to form a single-species superfluid~\cite{supp}. 
Semiclassically, this phase separation becomes energetically favored when $t_B / t_A > \left( 1 - f \right) / f$~\cite{supp}.  
We indeed find this phase-separated region using iDMRG, with a crossover  that follows the expected semiclassical trend, as shown in Fig.~\ref{fig:phasediagram}(b).
There is a first-order transition between the superfluid and phase-separated regions at finite $t_B/t_A$; meanwhile, the transition between supersolidity and phase separation is difficult to probe numerically, though the supersolid seems to be unstable to values of $t_B/t_A$ even slightly different from unity.

Three remarks are in order.
First, we note that the preparation of a supersolid phase, which combines the dissipationless flow of a superfluid with the periodic spatial density modulations of a solid~\cite{boninsegni2012colloquium}, has been the focus of much intense theoretical~\cite{henkel2010three,cinti2010supersolid,kora2019patterned} and experimental~\cite{bottcher2020new} effort beginning with the original proposal of solid $^4$He~\cite{balibar2010enigma}.
While initial observations in ultracold atomic systems were limited to systems where the structure formation was imposed by external light~\cite{li2017stripe,leonard2017supersolid,leonard2017monitoring,morales2018coupling}, more recent experiments have demonstrated the formation of metastable supersolids in which dipolar quantum droplets self-assemble into regular arrays~\cite{tanzi2019observation,bottcher2019transient,chomaz2019long}.
However, aided by the long coherence times of the Rydberg platform, our current proposal should facilitate the preparation of long-lived \textit{lattice} supersolids. 

Second, we note that lattice symmetry breaking (i.e., solidity) and XY long-range order (i.e., superfluidity) have been separately observed in spin-1/2 Rydberg atom array experiments~\cite{ebadiQuantumPhasesMatter2021,schollQuantumSimulation2D2021,chenContinuousSymmetryBreaking2023}; their combined manifestation as a supersolid phase, in a purely hardcore hopping model, is a novel feature of our three-level scheme.
This is made possible by the $\mathrm{U}(1)\times\mathrm{U}(1)$ symmetry, i.e., that there are two distinct bosonic species which may occupy separate regions of the lattice.
In comparison, we find that the ground state of $H_{\rm XY}$ (which has only one conserved charge, $S^z_\textrm{tot}$) is a superfluid irrespective of the filling fraction~\cite{supp}.
In such cases (and even in spin-1/2 systems), a supersolid phase can also be achieved with direct density-density interactions~\cite{wang:2009}.

Finally, we note that the existence of the supersolid phase is a consequence of the long-ranged, antiferromagnetic nature of our model, where frustration from the long-range tails favors the formation of supersolid stripes.
Indeed, neither the dipolar ferromagnetic model nor the nearest-neighbor model exhibit a supersolid ground state~\cite{supp}.

\begin{figure}%[H]
    \centering
\includegraphics[width=\columnwidth]{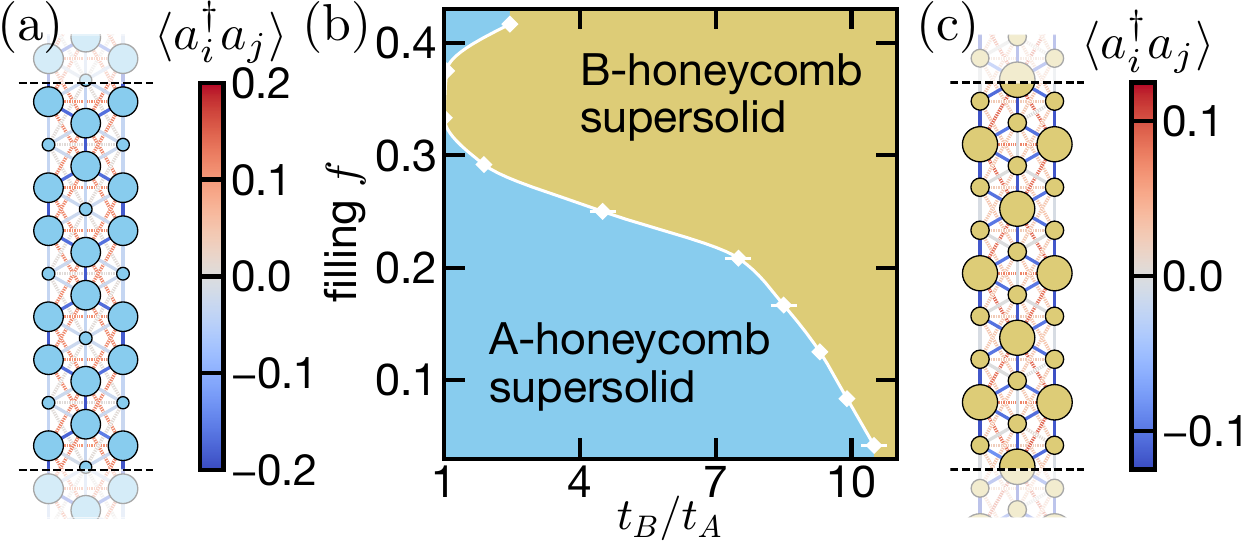}
    \caption{Phase diagram (b) of $H_{\rm AB}$ on the triangular lattice as a function of the filling $f$ and the hopping ratio $t_B/t_A$.  Representative real-space plots are shown for the two supersolid phases observed: (a) A-honeycomb  and (c) B-honeycomb, with the circle size denoting the onsite $A$ boson density, and the bond color scale denoting off-diagonal correlations.}
    \label{fig:triangular}
\end{figure}

The fragility of the square lattice supersolid to small changes in interaction strengths is caused by the weakness of the frustration.
To stabilize the supersolid phase, we now introduce additional \textit{geometric} frustration and study $H_{\rm AB}$ and $H_{\rm XY}$ on the triangular lattice.
For $H_{\rm AB}$, we again focus on symmetric fillings $f_A=f_B=f$.

Similarly to the square lattice, $H_{\rm XY}$ also exhibits superfluid (120\textdegree) order on the triangular lattice. 
By contrast, for $H_{\rm AB}$, we find a remarkably stable supersolid phase in all regimes studied, in both filling (down to $f=1/12$) and interaction strength (up to $t_B/t_A=10$) [Fig.~\ref{fig:triangular}(b)].
The triangular lattice supersolid is similar to that of the square lattice insofar as the $A$ and $B$ bosons separate and the superfluid order parameter $| \langle a_i^\dagger a_j e^{i \mathbf{k} \cdot \mathbf{r}_{ij}} \rangle |$ simultaneously peaks near the corners of the Brillouin zone. 
However, the nature of the translation symmetry breaking is distinct, and the hole density is also slightly nonuniform.

The triangular-lattice supersolid arises from a spontaneous separation of the $A$ and $B$ bosons into the honeycomb [Fig.~\ref{fig:triangular}(a)] and the intercalated triangular [Fig.~\ref{fig:triangular}(c)] sublattices of the base triangular lattice.
One species of bosons localizes on the bipartite honeycomb sublattice, where it forms a single-component superfluid, which is only weakly frustrated due to the long-range couplings.
The other species preferentially occupies the triangular sublattice while maintaining a lower nonzero density on the honeycomb sublattice. 
Each triangular sublattice site is anticorrelated with its hexagonal sublattice neighbors (i.e.,~$\langle b_i^\dagger b_j \rangle <0$), and different sites of the triangular sublattice are positively correlated.
% this above sentence is cuttable if we need to cut. 
Overall, the triangular-lattice supersolid phase breaks both U(1) symmetries, translation, and $A$-$B$ exchange symmetry (when $t_A=t_B$).

The preferential separation of the two bosonic species into the honeycomb and triangular sublattices persists for all filling fractions and interaction strengths.
For $t_A \neq t_B$, the boson species are no longer interchangeable: in Fig.~\ref{fig:triangular}(b), we show a phase diagram of the more energetically favorable choice between the $A$ bosons populating the triangular sublattice and the $B$ bosons populating the honeycomb sublattice, or vice versa.

\begin{figure}%[H]
    \centering
    \includegraphics[width=1.0\columnwidth]{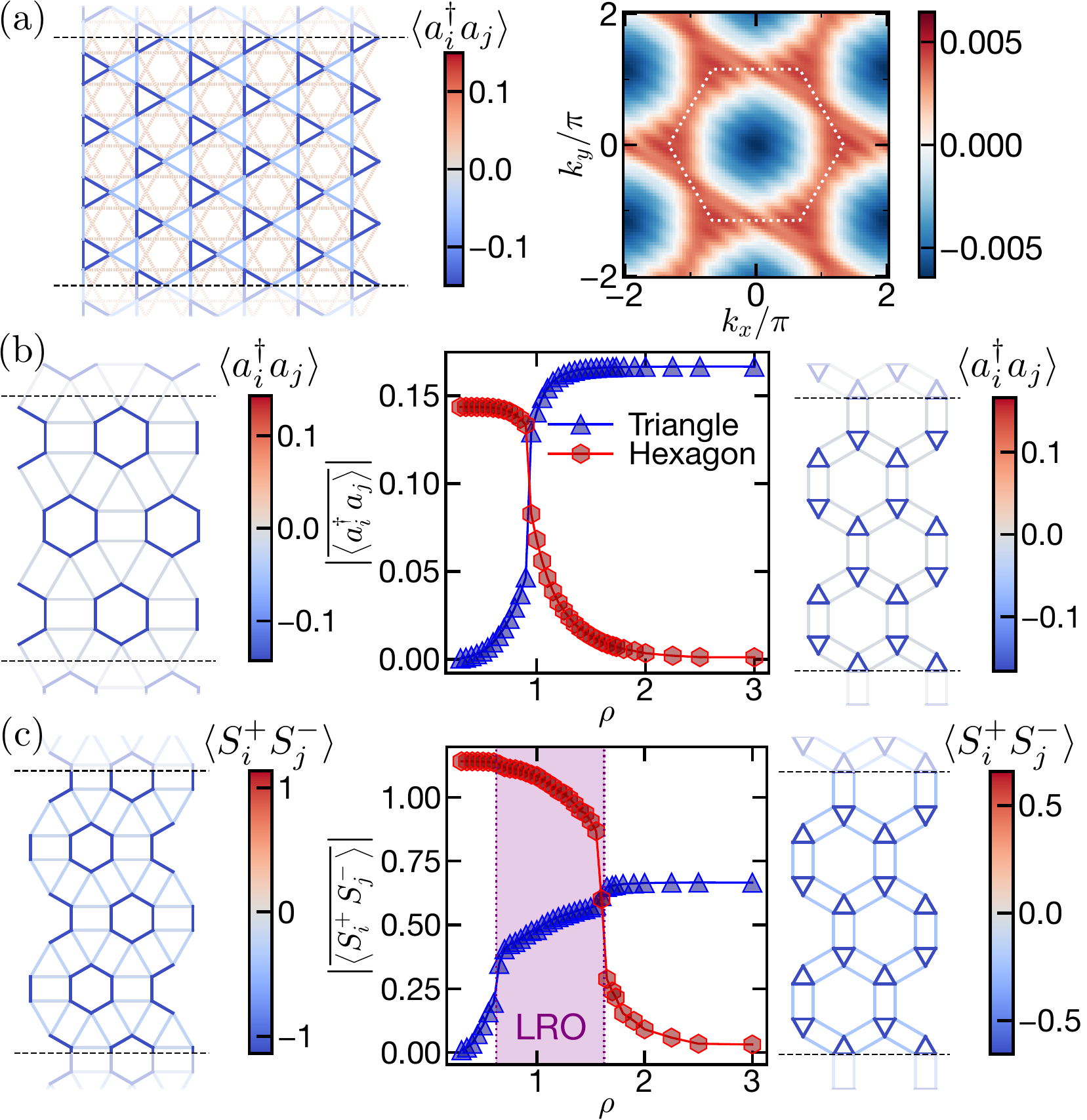}
    \caption{(a) Simplex solid on the kagome lattice, with $f=1/3, t_B/t_A=1$. All onsite densities are uniform, while correlations $\langle a_i a_j^\dagger \rangle$ are shown to be stronger (more negative) on the simplex triangles, as plotted on the left. The structure factor $\sum_{ij} \langle a_i^\dagger a_j e^{i \mathbf{k} \cdot \mathbf{r}_{ij}} \rangle$ (right) indicates that there is no long-range order, meaning simplices do not strongly interact. (b) Ground states of $H_{\rm AB}$ on the ruby lattice for varying aspect ratios $\rho$, with filling $f=1/3$. A ``hexagon paramagnet'' is formed at low $\rho$, with uniform boson densities and strongly anticorrelated hexagon bonds. The transition to the triangle-simplex paramagnet is first order. (c) Analogous ground states for $H_{\rm XY}$ showing a hexagon paramagnet, a long-range-ordered state, and a triangle-simplex paramagnet, all separated by first-order phase transitions.}
    \label{fig:simplex}
\end{figure}

%%%%%%%%%%%%%%%%%%%%%%%%%%%%%%%%
% SIMPLEX PHASES
%%%%%%%%%%%%%%%%%%%%%%%%%%%%%%%%
\emph{Simplex phases}.---Akin to the triangular lattice, the kagome lattice is a natural choice to investigate phases of frustrated antiferromagnets.
However, we find that the ground-state physics on the kagome lattice is profoundly different.
Unlike the square and triangular lattices, neither the $H_{\rm AB}$ nor $H_{\rm XY}$ ground states feature superfluid order or translation symmetry breaking.
Instead, their ground states are each characterized by a pattern of strong local correlations [Fig.~\ref{fig:simplex}(a)] in which spins on half of the corner-sharing triangles strongly trimerize.
That is, the three spins form an approximate \textit{simplex state},
\begin{equation}
    \left| \Delta \right> = \sum_{ijk \in \left\{ -1, 0, 1 \right\}} \epsilon_{ijk} \left \vert i,j,k \right \rangle,
\end{equation}
where $\epsilon_{ijk}$ is the totally antisymmetric Levi-Civita symbol.
Crucially, $\left| \Delta \right>$ minimizes both $H_{\rm AB}$ and $H_{\rm XY}$ on a single triangle, and can be tiled across the whole kagome plane by selecting half the lattice triangles (i.e., the rightward ones) as local simplices. 
Doing so spontaneously breaks spatial reflection, inversion, and $C_6$ rotational symmetries, and hence represents a simplex solid phase~\cite{asakawaPerturbationalExpansionAntiferromagnetic1993,arovasSimplexSolidStates2008,hermeleTopologicalLiquidsValence2011}.
We note that this phase was previously reported as the ground state of an $\mathrm{SU}(3)$-symmetric spin-1 Heisenberg model on the kagome lattice~\cite{corbozSimplexSolidsSU2012,changlaniTrimerizedGroundState2015,xuPhaseDiagramChiral2023}.
Our results show such an enlarged symmetry is not strictly necessary, opening the door to more realistic experimental investigations.

Spin trimerization can also occur on other lattices with triangular motifs---for instance, ruby lattices, in which the nearest-neighbor bonds belong to either individual triangles or hexagons [Fig.~\ref{fig:simplex}(b,c)]. 
Tuning the ratio $\rho$ between the hexagonal and triangular bond lengths modulates this lattice geometry between two limits: when $\rho\gg 1$, the lattice hosts small, strongly interacting triangles, while in the $\rho\ll 1$ limit, it features small, strongly interacting hexagons.
In the $\rho\gg 1$ limit one thus expects a simplex phase (for both $H_{\rm AB}$ and $H_{\rm XY}$), but this state should be destabilized as $\rho$ decreases and the hexagonal interactions become dominant.

The ground states of $H_{\rm AB}$ (setting $f=1/3$) and $H_{\rm XY}$ as a function of $\rho$ are depicted in Fig.~\ref{fig:simplex}(b,c).
As expected, both Hamiltonians realize a simplex solid over the triangular bonds of the ruby lattice for large $\rho$. 
Unlike the kagome case, this simplex ``solid'' does not break any spatial symmetries---it can be smoothly deformed to a trivial paramagnet without undergoing any phase transition.
% CHECK THIS!!!!!
%\MJB{TODO: What is the rep of $C_3$ on $\ket{\Delta}$?}
% (VL: it's the trivial rep, since Levi-Cevita is cyclic)
%
For smaller $\rho$, both models instead exhibit strongly anticorrelated hexagonal bonds.
For $H_{\rm AB}$, the two phases are separated by a first-order phase transition [Fig.~\ref{fig:simplex}(b)], while for $H_{\rm XY}$, there is an intervening long-range-ordered phase (superfluid) with first-order phase transitions on both sides [Fig.~\ref{fig:simplex}(c)].
Finally, we note that, for $H_{\rm AB}$,  simplex physics can also coexist with symmetry-breaking density patterns when $f \neq 1/3$, in both the kagome and ruby lattices~\cite{supp}.

\emph{Experimental considerations}.---Our hardcore hopping Hamiltonian, $H_{\rm AB}$, is naturally realized in optical tweezer arrays of alkali or alkaline earth atoms, or alternatively by using the rotational states of dipolar molecules~\cite{gorshkovQuantumMagnetismPolar2011}.
%
%In the context of Rydberg systems, experiments utilizing multiple Rydberg levels have also been recently demonstrated~\cite{Kanungo2022}.
%
While the majority of ground-state phases that we predict exist for a wide range of  $t_B/t_A$, 
the square-lattice supersolid seems stable only near  $t_B/t_A=1$; this symmetric case is also needed for realizing the exact form of $H_{\rm XY}$. 
Fortunately, it is relatively straightforward to find triplets of Rydberg levels with nearly symmetric hopping strengths.
% check this!!
Taking $^{87}\textrm{Rb}$ as an example, for sufficiently large principle quantum numbers $n\gtrsim 50$, one can
choose $\ket{\mathsf{S}_B} = \ket{n \; \mathsf{S} \; m_J}$, $\ket{\mathsf{P}_0} = \ket{n \; \mathsf{P} \; \frac{1}{2}}$, $\ket{\mathsf{S}_A} = \ket{(n+1) \; \mathsf{S} \; m_J}$,  and $m_J = \pm \frac 1 2$, which yields  $\abs{t_A-t_B}/t_A \lesssim  10\%$.
% check with max and lysander
%
To realize $H_{\rm XY}$, one additionally needs to match the detunings, $\delta_A \approx \delta_B$, which can be achieved by utilizing an electric field to 
tune to a F\"orster resonance~\cite{ravetsCoherentDipoleDipole2014}.
%
%Incidentally, this choice of levels also results in $\sim 10\%$ difference in hopping strength.
%
We note that the essential physics of the ground states we find for $H_{\rm XY}$ derive from its U(1) symmetry, which is realized solely by matching the detunings, and should thus be insensitive to small differences in the values of $t_A$ and $t_B$, as well as to additional $S^zS^z$ terms that may appear as a result of the applied field.
%
% We also note that the van der Waals interactions between Rydberg states also scale as $1/r^3$ in the vicinity of a F\"orster resonance, which may lead to additional density-density interactions in the U(1) case.
%

In principle, the supersolid and simplex phases we observe can be adiabatically prepared by: (i) initializing the system in a $z$-basis product state accompanied by local detunings, and then (ii) ramping these detunings to zero~\cite{chenContinuousSymmetryBreaking2023}.
Since a local detuning does not break either the 
$\mathrm{U}(1)\times\mathrm{U}(1)$ symmetry of $H_\textrm{AB}$ or the $\mathrm{{\rm U}(1)}$ symmetry of $H_\textrm{XY}$, one can directly tune $f_A$ and $f_B$ by initializing  product states with varying filling fractions.

%%%%%%%%%%%%%%%%%%%%%%%%%%%%%%%%%%%%%%%%%%%%%%%%%%%%
% CONCLUSION/OUTLOOK/DISCUSSION
%%%%%%%%%%%%%%%%%%%%%%%%%%%%%%%%%%%%%%%%%%%%%%%%%%%%

In conclusion, our work provides a roadmap for Rydberg array experiments aiming to probe correlated phases beyond the spin-1/2 setting.
Looking forward, a number of intriguing questions emerge. 
First, how does phase separation arise dynamically~\cite{groverQuantumDisentangledLiquids2014,huse2024},  either during adiabatic state preparation on in response to a global quench?  
Second, building on recent conjectures for the connection between continuous symmetry breaking and quantum sensing~\cite{perlinSpinSqueezingShortRange2020,blockScalableSpinSqueezing2024,comparin}, is it possible to prepare metrologically useful entanglement via the dynamics of $H_\textrm{AB}$ and $H_\textrm{XY}$?
Finally, by analogy to the melting of quantum dimers in spin-1/2 antiferromagnets, are there strategies for  distilling a topological spin-1 liquid~\cite{greiterNonAbelianStatisticsQuantum2009,groverNonAbelianSpinLiquid2011,liTopologyCriticalityResonating2014} from our simplex solids?

\textit{Note added.}---During completion of this work we became aware of another proposal to study supersolidity with Rydberg tweezer arrays, utilizing a two-level scheme with both dipolar exchange and van der Waals interactions~\cite{homeierSupersolidityRydbergTweezer2024}.

\textit{Acknowledgements}---We gratefully acknowledge the insights of and discussions with G. Bornet, A. Browaeys, L. Christakis, J. Hauschild, L. Homeier,  T, Lahaye, and L. Pollet.
This work was supported by the U.S. Department of Energy, Office of Science, National Quantum Information Science Research Centers, Quantum Systems Accelerator (QSA) and the Harvard-MIT Center for Ultracold Atoms.
R.S. is supported by the Princeton Quantum Initiative Fellowship. 
J.K. acknowledges support from the Air Force Office of Scientific Research through the MURI program (FA9550-21-1-0069).
N.Y.Y. acknowledges support from a Simons Investigator award.

%%%%%%%%%%%%%%%%%%%%%%%%%%%%%%%%%%%%%%%%%%%%%%%%%%%%
% Acknowledgments
%%%%%%%%%%%%%%%%%%%%%%%%%%%%%%%%%%%%%%%%%%%%%%%%%%%%

%%%%%%%%%%%%%%%%%%%%%%%%%%%%%%%%%%%%%%%%%%%%%%%%%%%%
% APPENDIX
% \clearpage % remove this at the end
% \section*{Appendix}

% \appendix

\bibliography{refs,zotero_refs,footnotes_and_other}

\end{document}

\title{Supplemental Material for ``Supersolidity and Simplex Phases in Spin-1 Rydberg Atom Arrays''}

\newcommand{\HarvardPhysicsAddress}{Department of Physics, Harvard University, Cambridge, Massachusetts 02138, USA}
\newcommand{\PrincetonAAddress}{Department of Physics, Princeton University, Princeton, NJ 08544, USA}
\newcommand{\PrincetonBAddress}{Princeton Center for Theoretical Science, Princeton University, Princeton, NJ 08544, USA}

\author{Vincent S. Liu}
\affiliation{\HarvardPhysicsAddress}

\author{Marcus Bintz}
\affiliation{\HarvardPhysicsAddress}

\author{Maxwell Block}
\affiliation{\HarvardPhysicsAddress}

\author{Rhine Samajdar}
\affiliation{\PrincetonAAddress}
\affiliation{\PrincetonBAddress}

\author{Jack Kemp}
\affiliation{\HarvardPhysicsAddress}

\author{Norman Y. Yao}
\affiliation{\HarvardPhysicsAddress}

\date{\today}
\maketitle

\section{Numerical Details: Cylinder $\mathbf{i}$DMRG}

\begin{figure*}
    \includegraphics[width=0.55\textwidth]{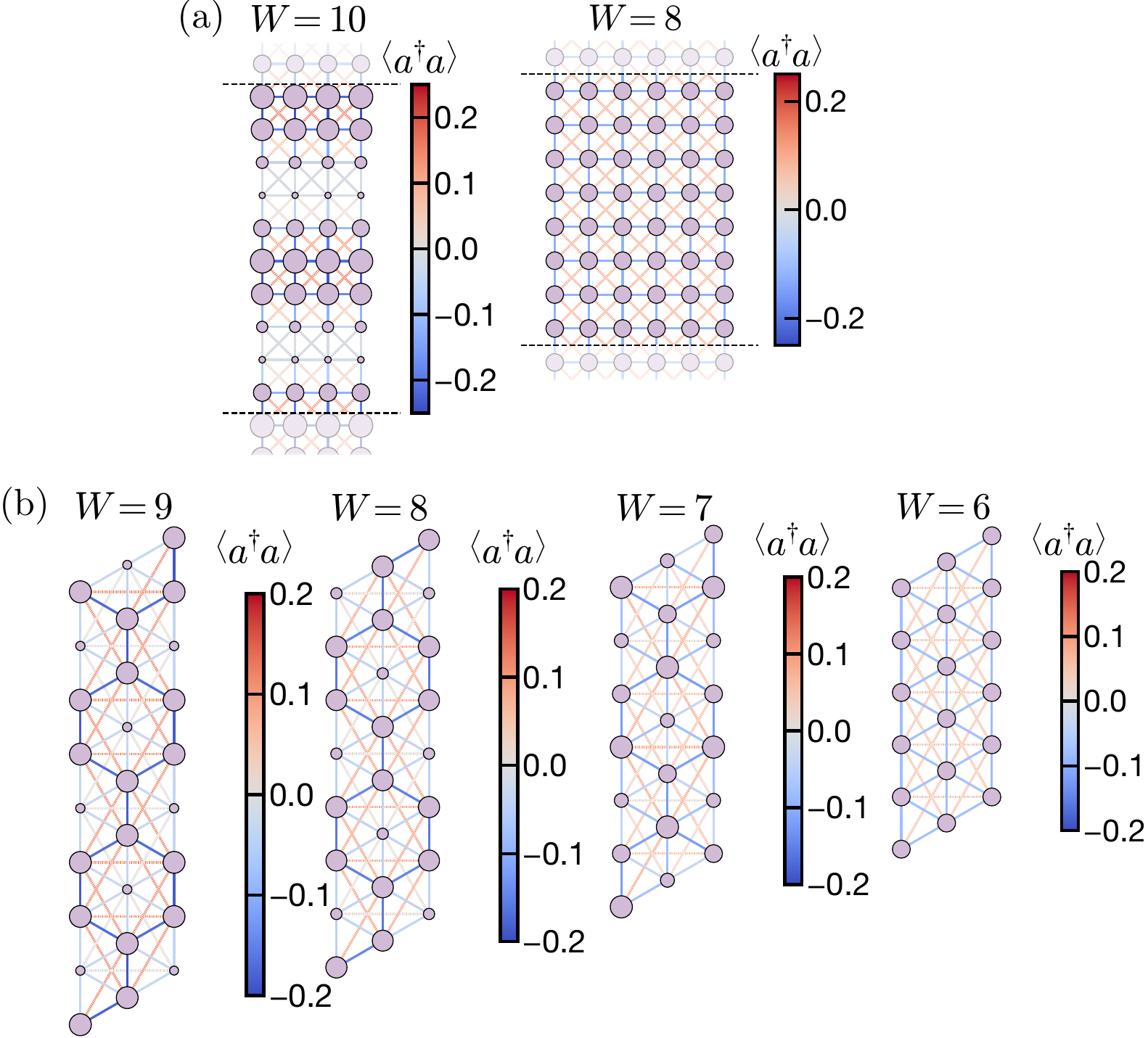}
    \caption{\label{fig:SI_widths} Real-space plots of the ground states of $H_\textrm{AB}$ on the (a) square and (b) triangular lattices, with higher values of $W$ showing supersolid order and lower values of $W$ showing spatially uniform superfluid order. For all plots, the circle size denotes the onsite $A$ boson density $\langle N_A \rangle$, while the bond color scale represents the values of the off-diagonal correlations $\langle a_i^\dagger a_j \rangle$. (a) The left plot with $W=10$ is at filling $f=0.3$, while the right plot with $W=8$ is at filling $f=1/3$, both of which are in the superfluid phase (main text, Fig.~2(b)). (b) All plots have filling $f=1/3$; interestingly, the $W=7$ cylinder produces a different supersolid than the $W>7$ geometries.
    }
\end{figure*}

We first review the application of cylinder iDMRG to our system.
Our simulations are performed using the open-source package TeNPy~\cite{tenpy} developed by Johannes Hauschild and Frank Pollmann.
As DMRG is a 1D algorithm, we represent our two-dimensional geometries as quasi-one-dimensional infinite cylinders and map the resulting lattice onto an infinite chain with longer-range interactions.
In particular, the finite cylinder circumference $W$ (typically six to twelve lattice sites) has important implications for the numerically determined ground state.
As shown in Fig.~\ref{fig:SI_widths}, both the square- and triangular-lattice supersolid phases of $H_\textrm{AB}$ are only obtained for higher cylinder widths, while DMRG results for lower cylinder widths instead remain in the superfluid phase.
Thus, it is essential to choose sufficiently wide cylinders to accurately represent the many-body quantum phases that most accurately describe the thermodynamic limit.
For phases that break translation symmetry, the choice of a translation-invariant ansatz (i.e., the length of the matrix product state along the infinite direction) is also important, as some system sizes are incommensurate with the desired ground state and thus fail to properly represent it.
To address this issue, we run our simulations on many cylindrical geometries, choosing the state with lowest energy density if the results differ.

\begin{figure*}
    \includegraphics[width=0.4\textwidth]{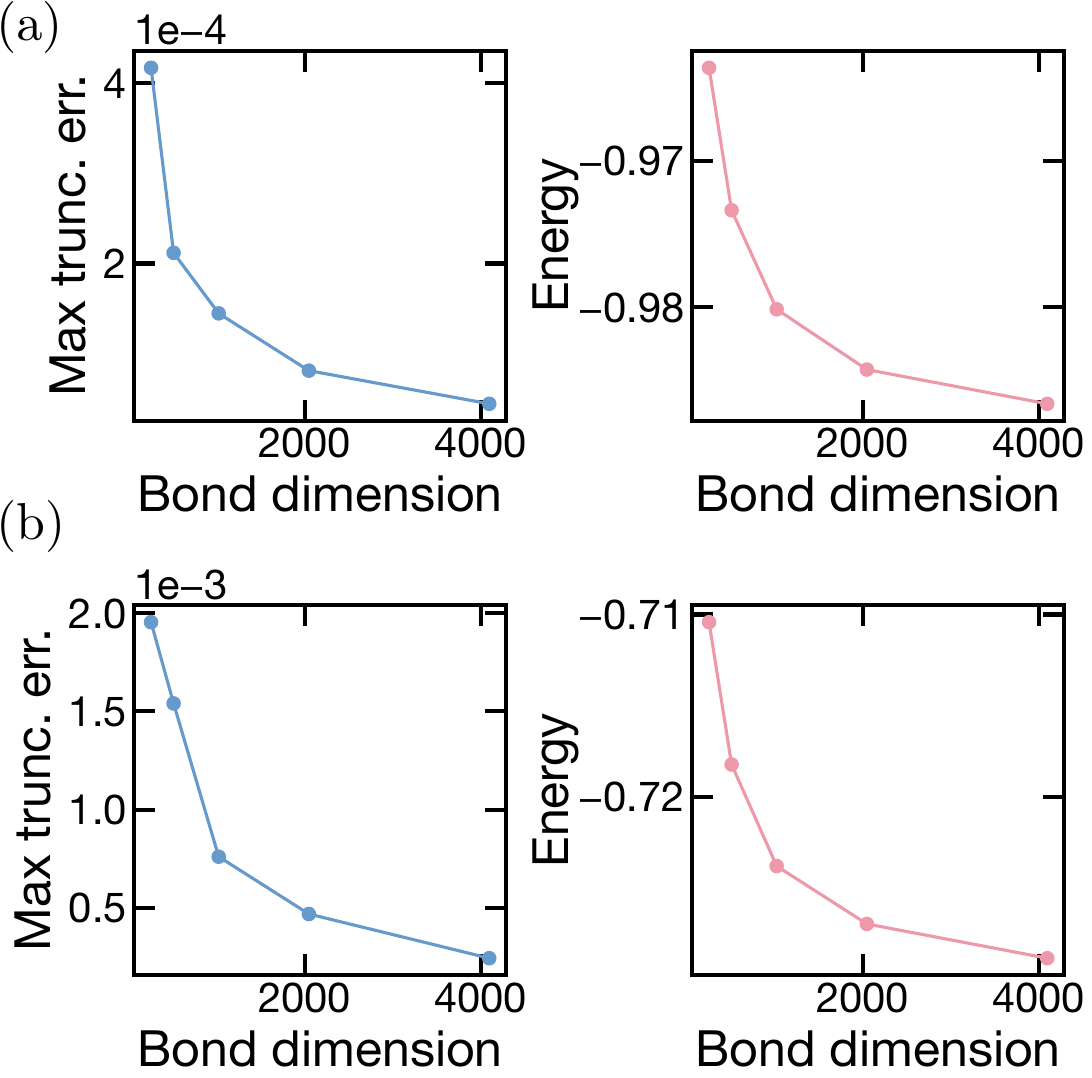}
    \caption{\label{fig:SI_convergence} Convergence of representative (a) supersolid  and (b) simplex phases of $H_\textrm{AB}$ as a function of bond dimension $\chi$, from $\chi=256$ to $\chi=4096$. Left: truncation error (norm of all truncated Schmidt values), right: calculated ground state energy.
    }
\end{figure*}

Apart from geometrical considerations, our DMRG studies make several important approximations.
First, the power-law dipolar interactions in our Hamiltonians $H_\textrm{AB}$ and $H^{}_{\rm{XY}}$ cannot be exactly represented as a matrix product operator (MPO).
We therefore truncate all interactions to zero beyond a maximum distance $d$.
We find that the choice of $d$ has no significant effect on the resulting phases beyond $d \sim 4$.
Second, our computations are necessarily restricted to finite bond dimensions.
In Fig.~\ref{fig:SI_convergence}, we show the effects of increasing the bond dimension $\chi$ on the calculated ground state for representative supersolid and simplex phases, with most measures converging by $\chi \sim 4096$.

We take multiple additional measures to improve the robustness of our iDMRG results and reduce the likelihood of getting trapped in local minima.
First, we increment the bond dimension, usually starting from $\chi \sim 256$ until our final value of $\chi \sim 4096$, multiple times per simulation.
Each time the bond dimension is increased, we activate the two-site DMRG ``mixer'', which performs subspace expansion to potentially break free from local minima.
Additionally, we initialize our simulations with multiple trial wavefunctions to better explore the landscape of phases.
These include uniform states and multiple random states, as well as trial wavefunctions informed by the conjectured ground state phases---for example, density patterns commensurate with the hypothesized symmetry breaking, or superfluid and supersolid states from previous DMRG simulations at different parameters.
In particular, our triangular lattice phase diagram in Fig.~3(b) of the main text  is constructed by seeding each point in parameter space with ans\"{a}tze of both the A-honeycomb and B-honeycomb supersolid phases.

\section{Additional results}

\subsection{Characterization of superfluid and supersolid phases}

\begin{figure*}
    \includegraphics[width=0.6\textwidth]{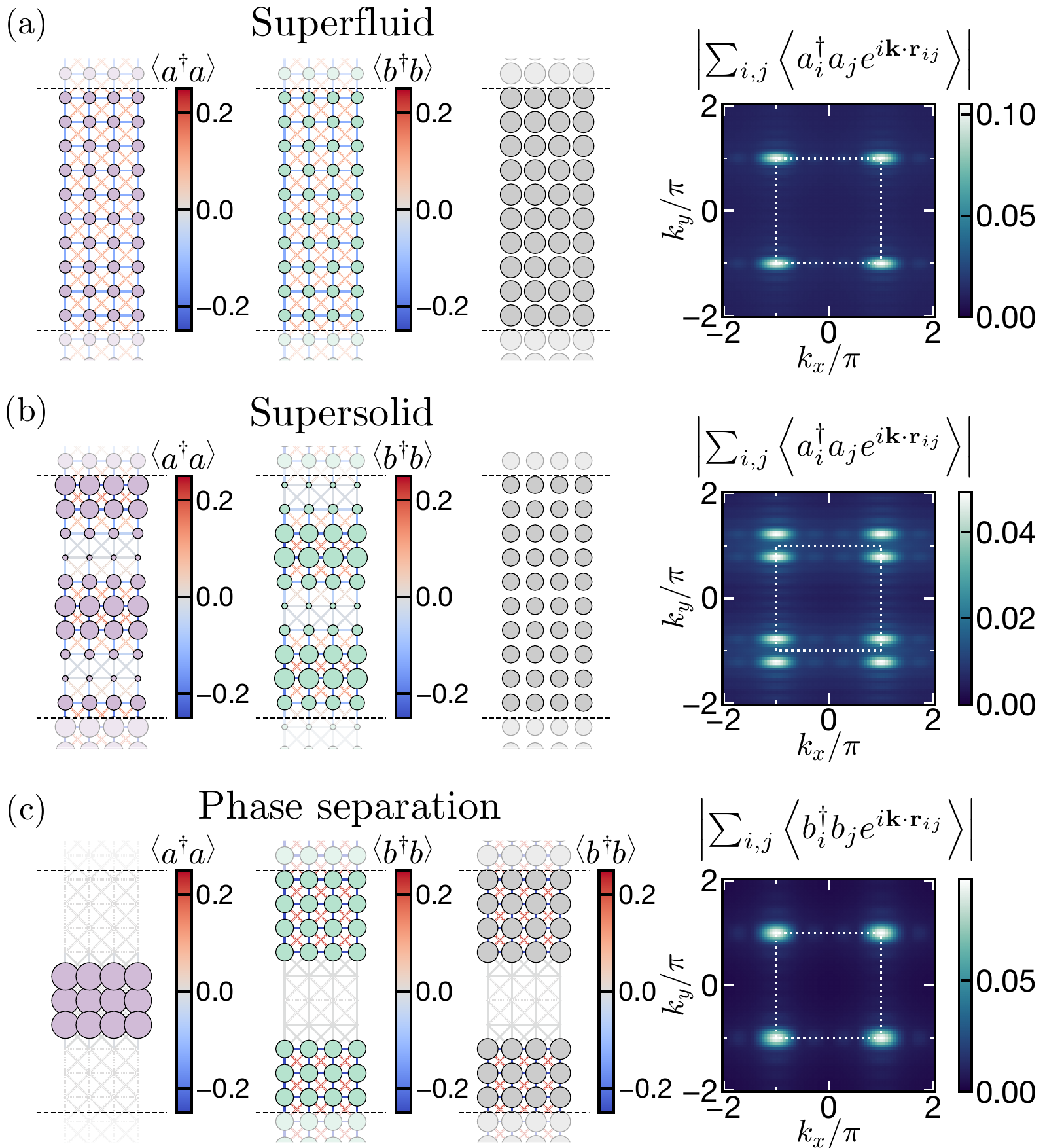}
    \caption{\label{fig:SI_sq_phases} Phases of $H_\textrm{AB}$ on the square lattice, with real-space plots of $A$ boson density, $B$ boson density, and hole density, from left to right, and the superfluid structure factor on the right. (a) Superfluid phase ($f=0.2$), with uniform real-space profiles and single peaks in the structure factor. (b) Supersolid phase ($f=0.3$), with opposing stripes of $A$ and $B$ boson density, uniform hole density, and a doubly peaked structure factor. (c) Phase separation ($f=0.3$, $t_B/t_A=100$), with $B$ bosons and holes forming a single-component superfluid.
    }
\end{figure*}

We now review the superfluid and supersolid phases of $H_\textrm{AB}$ in greater detail, as mentioned in the main text.
Starting with the square lattice, the full real-space density profiles of the supersolid, stripe supersolid, and phase-separated regimes are shown in Fig.~\ref{fig:SI_sq_phases}.
Both the superfluid and supersolid phases have uniform hole density, even when the supersolid phase breaks the translation symmetry of both $A$ and $B$ bosons.
Indeed, in the supersolid phase, the two bosonic species form opposing stripes to keep the hole density uniform.
In this phase, we also find that there is long-range off-diagonal order in both the infinite and periodic directions, with $\langle a^\dagger_i a_j \rangle$ correlations being large even across different stripes.
In the presence of phase separation (Fig.~\ref{fig:SI_sq_phases}(c)), the hole density is nonuniform, with the $A$ bosons having precipitated out so that the $B$ bosons and holes can form a superfluid.

\begin{figure*}
    \includegraphics[width=0.75\textwidth]{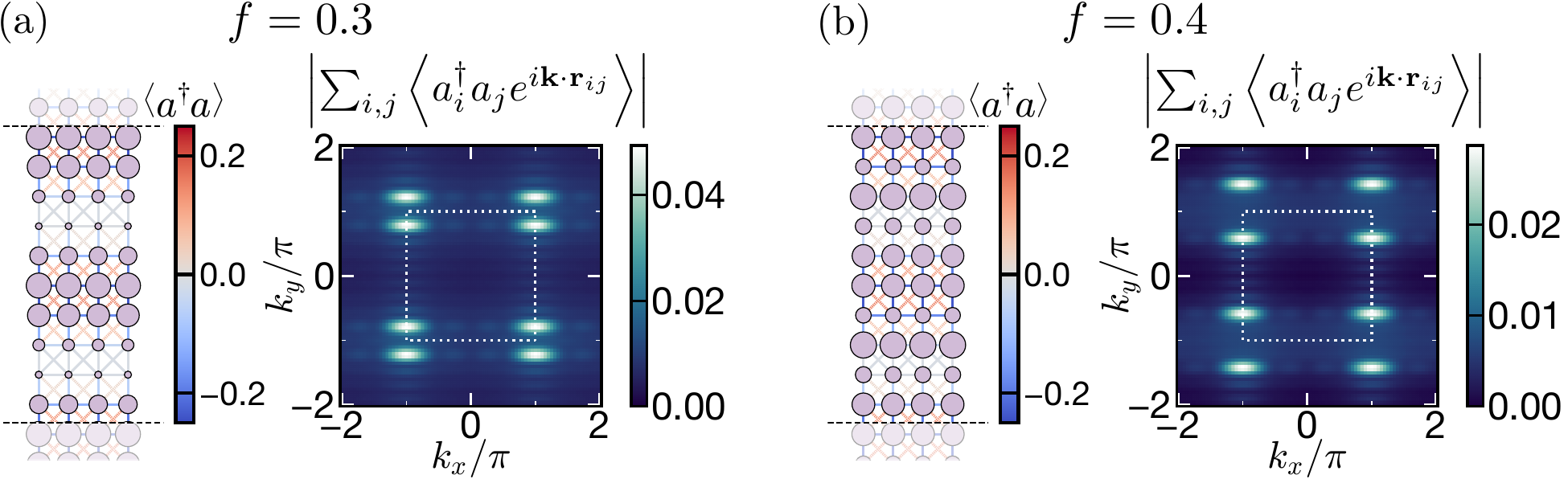}
    \caption{\label{fig:SI_sq_ss2} Comparison of square supersolid ground states of $H_\textrm{AB}$  for fillings (a) $f=0.3$ and (b) $f=0.4$, with real-space plots of $A$ boson density and off-diagonal $\langle a^\dagger_i a_j \rangle$ correlations shown on the left and  the structure factors plotted on the right. The $f=0.4$ supersolid exhibits narrower stripes and a larger peak separation in the Brillouin zone.
    }
\end{figure*}

The behavior of the supersolid phase at different fillings merits further explanation.
As the filling $f$ is increased above $1/4$, the real-space supersolid stripes become smaller and the reciprocal-lattice peaks become further separated.
This is shown in Fig.~\ref{fig:SI_sq_ss2}, which compares the superfluid at $f=0.3$ and $f=0.4$.
Due to the finite cylinder width $W=10$ in our simulations, the width of the supersolid stripes is difficult to exactly discern; however, we estimate that, with peaks in structure factor at $\left( \pm \pi, \pm \pi \pm \Delta k \right)$, density oscillations are proportional to $\sin \left( 2 \Delta k y \right)$.
We hypothesize that, in the thermodynamic limit, $f=1/4$ represents a transition from the superfluid phase to a supersolid with infinitely wide stripes, and that the  width of the stripes continuously decreases as a function of $f$ for $f>1/4$.
Testing this prediction is beyond the capabilities of iDMRG but is a potential topic for further study by variational methods.

\begin{figure*}
    \includegraphics[width=0.9\textwidth]{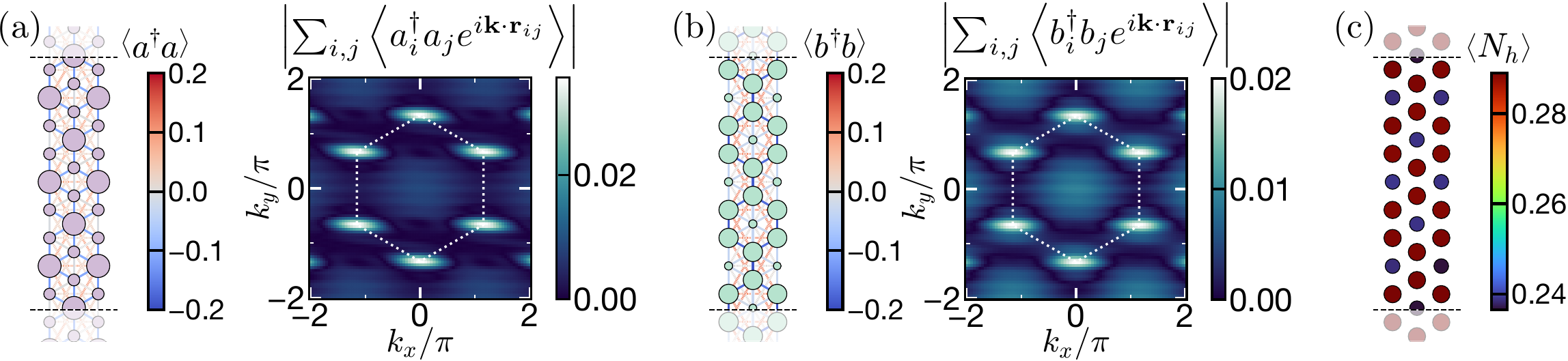}
    \caption{\label{fig:SI_tri_phases} Representative ground states of $H_\textrm{AB}$ on the triangular lattice (with $f=1/3$). (a) $A$ bosons forming the intercalated triangular-lattice supersolid, with the structure factor indicating $120^\circ$ long-range order. Left: real-space plot of onsite $A$ boson density (circle sizes) and $\langle a^\dagger_i a_j \rangle$ correlations (bond color scale); right: structure factor $\left| \sum_{i,j} \left< a^\dagger_i a_j e^{i \mathbf{k} \cdot \mathbf{r}_{ij}} \right> \right|$. (b) $B$ bosons forming the honeycomb supersolid, with peaks at the corners of the honeycomb Brillouin zone. Left: real-space plot of onsite $B$ boson density (circle sizes) and $\langle b^\dagger_i b_j \rangle$ correlations (bond color scale); right: structure factor $\left| \sum_{i,j} \left< b^\dagger_i b_j e^{i \mathbf{k} \cdot \mathbf{r}_{ij}} \right> \right|$.
    }
\end{figure*}

Moving to the triangular lattice supersolids, as shown in Fig.~\ref{fig:SI_tri_phases}, we remark that the hole density is nonuniform, unlike the square supersolid case, with different densities on the honeycomb sublattice and intercalated triangular sublattice.
The specific sublattice with higher density depends on the filling and parameter regime.
We also remark, as explained in the main text, that both the honeycomb and intercalated triangular supersolids exist in the same state, as the phases of the two respective boson species.

\subsection{Necessary conditions for supersolids}

\begin{figure*}
    \includegraphics[width=0.97\textwidth]{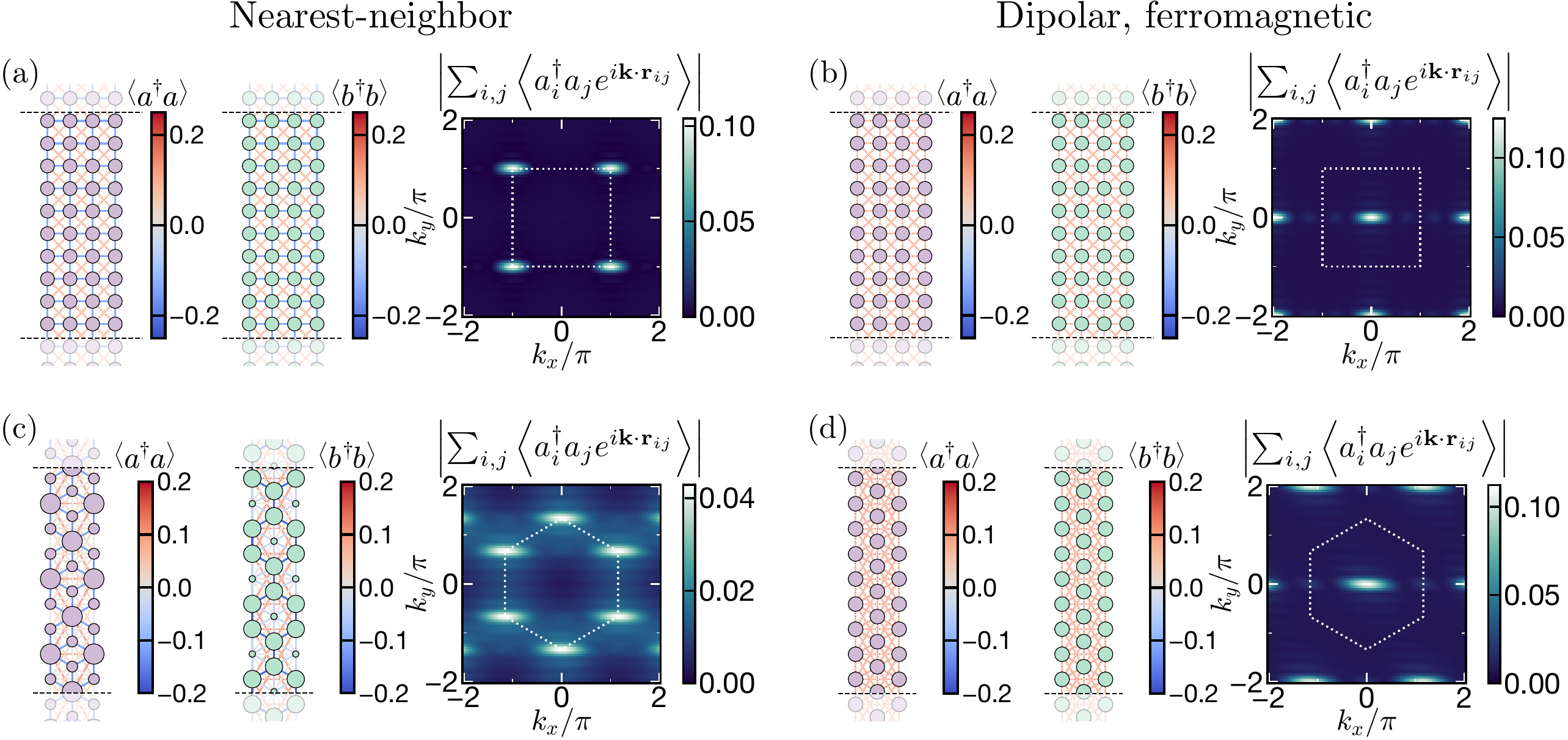}
    \caption{\label{fig:SI_nn_fm} All panels---Left: real-space plots of boson densities $\langle N_A \rangle$, $\langle N_B \rangle$ (circle sizes) and off-diagonal correlations $\langle a^\dagger_i a_j \rangle$, $\langle b^\dagger_i b_j \rangle$ (bond color scales). Right: structure factor $\left| \sum_{i,j} \left< a^\dagger_i a_j e^{i \mathbf{k} \cdot \mathbf{r}_{ij}} \right> \right|$. (a) Ground state of $H_\textrm{AB}$ ($f=0.3$) on the square lattice with only nearest-neighbor interactions, forming a superfluid phase. (b) Ground state of $H_\textrm{AB}$ ($f=1/3$) on the triangular lattice with only nearest-neighbor interactions, forming the honeycomb and intercalated triangular supersolids. (c,d) Ground states of $H_\textrm{AB}$ with $t_A,t_B=-1$ on the (c) square ($f=0.3$) and (d) triangular ($f=1/3$) lattices, showing uniform superfluids with positive correlations.
    }
\end{figure*}

We now turn to the conditions that enable the formation of ground-state supersolids in $H_\textrm{AB}$.
As mentioned in the main text, we believe that both the square-lattice and triangular-lattice supersolids emerge due to frustration---from either the long-range dipolar tails on the square lattice or the geometric frustration on the triangular lattice.
To test these hypotheses, we compare the ground state of the dipolar, antiferromagnetic ($t_A,t_B>0$) case we have previously explored to two alternatives: the case where only nearest-neighbor interactions are nonzero (removing any frustration from the long-range tail), and the case of ferromagnetic ($t_A,t_B<0$) interactions (removing all frustration).
The results are shown in Fig.~\ref{fig:SI_nn_fm}.
As expected, there are no supersolids in the ferromagnetic case, with both the square- and triangular-lattice ground states forming a superfluid.
In the antiferromagnetic case, the square lattice forms a superfluid due to the lack of frustration, while the triangular-lattice supersolid is qualitatively unaffected.

\subsection{Supersolids in finite geometries}

\begin{figure*}
    \includegraphics[width=\textwidth]{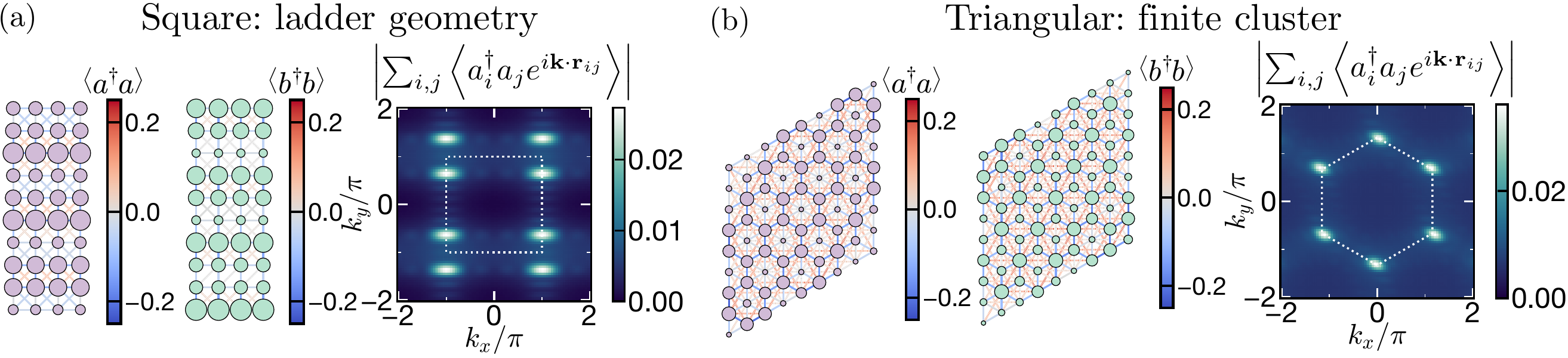}
    \caption{\label{fig:SI_finite} All panels---Left: real-space plots of boson densities $\langle N_A \rangle$, $\langle N_B \rangle$ (circle sizes) and off-diagonal correlations $\langle a^\dagger_i a_j \rangle$, $\langle b^\dagger_i b_j \rangle$ (bond color scales). Right: structure factor $\left| \sum_{i,j} \left< a^\dagger_i a_j e^{i \mathbf{k} \cdot \mathbf{r}_{ij}} \right> \right|$. (a) Ground state of $H_\textrm{AB}$ ($f=0.4$) on an infinite square-lattice ladder  geometry, exhibiting characteristic stripes and doubled peaks in the Brillouin zone. (b) Ground state of $H_\textrm{AB}$ ($f=1/6$) on a $9 \times 9$ open triangular-lattice  cluster, displaying sublattice separation and long-range order.
    }
\end{figure*}

To realize the above supersolid phases in near-term cold-atom experiments, they must remain stable in the presence of edge effects and on finite cluster geometries.
Here, we investigate both the square- and triangular-lattice supersolids under open boundary conditions.
For the triangular lattice, we require a width of at least $W=8$ to obtain the desired supersolid phase (Fig.~\ref{fig:SI_widths}).
Fortunately, finite DMRG is able to reliably converge the commensurate $9 \times 9$ triangular lattice cluster; as shown in Fig.~\ref{fig:SI_finite}(b), the separation of the $A$ and $B$ bosons into the honeycomb and intercalated triangular sublattices seems to persist, with long-range order also preserved---thus, we expect that the triangular-lattice ground state is stable to open geometries.
On the square lattice, the convergence of finite DMRG on the desired $10 \times 10$ cluster is poor; the boson densities are, however, nonuniform, a weak sign that supersolid order is attempting to form.
For a more rigorous investigation of the square-lattice supersolid on an open geometry, we instead use iDMRG with open boundary conditions in the $y$-direction (known as the ``ladder'' geometry).
Here, we see the characteristic supersolid stripes and doubled peaks in the structure factor, as shown in Fig.~\ref{fig:SI_finite}(a).
These preliminary results are promising for the experimental realization of these supersolid phases; a more careful investigation with quantum Monte Carlo methods, which could also reveal the stability of these phases at nonzero temperatures, may be a direction for further research.

\subsection{Semiclassical energetic analysis for phase separation}

Here, we review a semiclassical argument for the energetic favorability of phase separation on the square lattice as mentioned in Fig.~2 of the main text.
Given the Hamiltonian $H_\textrm{AB}$ with a fixed value of $t_B/t_A$ and filling $f$, we estimate the energies of two wavefunction ans\"atze: a uniform superfluid, and a completely phase-separated state, with all $A$ bosons precipitated out.
In the superfluid, all sites have densities $\langle N_A \rangle = \langle N_B \rangle = f$, $\langle N_h \rangle = 1-2f$.
We assume that the energy density $E$ is proportional to $\sum_{\eta \in \{ A,B \}} t_\eta \langle N_\eta \rangle \langle N_h \rangle$, as the hopping energy between $A$ and $B$ bosons and holes depends on the product of their densities.
Thus, in the superfluid, we have $E/t_A \sim ( 1+t_B/t_A ) f (1-2f)$.

In the phase-separated case, a proportion $f$ of all sites has $\langle N_A \rangle = 1$, $\langle N_B \rangle = \langle N_h \rangle = 0$, while a proportion $1-f$ of all sites has $\langle N_A \rangle = 0$, $\langle N_B \rangle = f/(1-f)$, $\langle N_h \rangle = (1-2f)/(1-f)$.
Thus, we have $E/t_A \sim t_B/t_A f (1-2f)/(1-f)$.
For phase separation to be energetically favorable, we must have $t_B/t_A f (1-2f)/(1-f) > ( 1+t_B/t_A ) f (1-2f)$, or equivalently,
\begin{align}
    \frac{t_B}{t_A} > \frac{1-f}{f}.
\end{align}
In this semiclassical treatment, the overall energy density $E$ is independent of the distribution of the boson densities $\langle N_A \rangle$ and $\langle N_B \rangle$ if the hole density $\langle N_h \rangle$ remains uniform.
Thus, the superfluid and the square-lattice supersolid cannot be energetically distinguished.

\subsection{Superfluid ground states of $H_{\rm{XY}}$}

\begin{figure*}
    \includegraphics[width=0.7\textwidth]{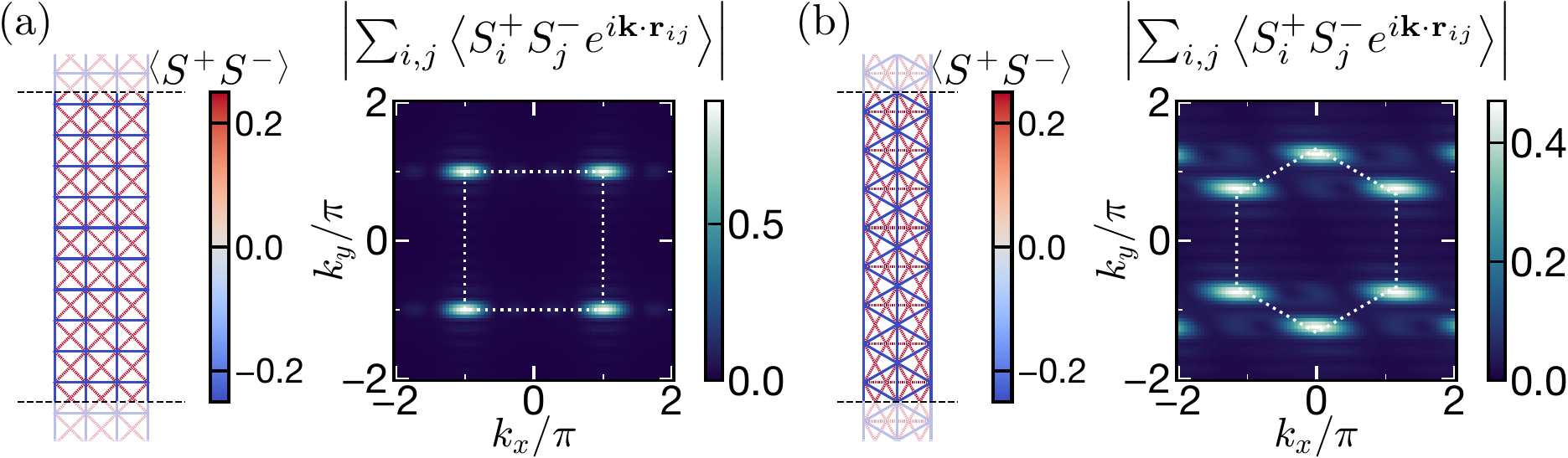}
    \caption{\label{fig:SI_xy} (a) Ground state of $H^{}_{\rm{XY}}$ on the square lattice with total magnetization $M_z=0$. (b) Ground state of $H^{}_{\rm{XY}}$ on the triangular lattice with total magnetization $M_z=0$. All panels---Left: real-space plots, with onsite $\langle S_z \rangle$ being uniformly zero and the bond color scale representing the off-diagonal spin correlator $\langle S^+_i S^-_j \rangle$. Right: spin structure factor $\left| \sum_{i,j} \left< S^+_i S^-_j e^{i \mathbf{k} \cdot \mathbf{r}_{ij}} \right> \right|$.
    }
\end{figure*}

As shown in Fig.~\ref{fig:SI_xy} and mentioned in the main text, $H^{}_{\rm{XY}}$ has a superfluid ground state in the regimes where $H_\textrm{AB}$ features ground-state supersolids.
In the language of spins, the square-lattice state exhibits XY long-range order and the triangular-lattice state exhibits $120^\circ$ order.
This aligns with the intuition that larger spins (spin-1) behave more classically than smaller spins (spin-1/2), given that it is possible that the ground state of the spin-1/2 $H^{}_{\rm{XY}}$ model on the triangular lattice hosts an interesting many-body state~\cite{Bintz_Liu_Hauschild_Khalifa_Chatterjee_Zaletel_Yao_2024}.

\subsection{Density patterns and simplex phases}

\begin{figure*}
    \includegraphics[width=0.3\textwidth]{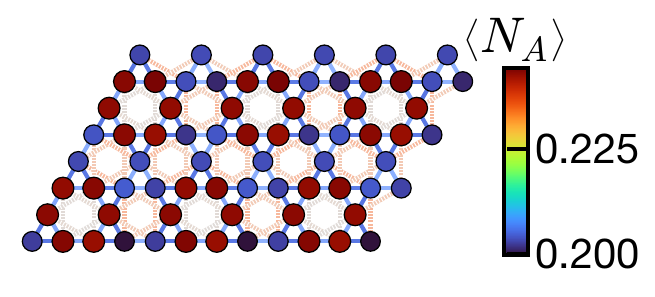}
    \caption{\label{fig:SI_simplex_density} Ground state of $H_\textrm{AB}$ on the kagome lattice for $f=1/4$, showing a symmetry-breaking density pattern. The colors of the circles represent onsite boson densities $\langle N_A \rangle$, while bond colors represent off-diagonal correlations $\langle a^\dagger_i a_j \rangle$.
    }
\end{figure*}

Turning briefly to simplices, we remark that symmetry-breaking density patterns can arise in place of or alongside the simplex physics described in the main text.
This is especially the case for fillings other than $f=1/3$ in $H_\textrm{AB}$, where the perfect simplex state $ \left| \Delta \right> = \sum_{ijk \in \left\{ -1, 0, 1 \right\}} \epsilon_{ijk} \left| i,j,k \right>$ cannot be naturally formed with equal fillings of $A$ and $B$ bosons and holes.
Figure~\ref{fig:SI_simplex_density} presents one such example on the kagome lattice.
The ruby lattice also shows similar physics, with nonuniform densities for $f \neq 1/3$, but is difficult to reliably converge with iDMRG.

\section{Derivation of the $t$-$J$-$V$-$W$ Hamiltonian from dressed interactions}

As discussed in the main text, dressing our three-level scheme with additional Rydberg levels provides a route to matching both detunings and transition dipole moments to help realize $H^{}_{\rm{XY}}$.
In fact, the additional interactions obtained via such dressing can also realize a bosonic $t$-$J$-$V$-$W$ Hamiltonian even without breaking the $\mathrm{{\rm U}(1)} \times \mathrm{{\rm U}(1)}$ symmetry of $H_\textrm{AB}$, which may represent an interesting direction in and of itself.
For comparison, the fermionic $t$-$J$-$V$-$W$ model has been explored in 1D, giving rise to interesting many-body states such as Luttinger liquids and topological superconductors~\cite{fazziniInteractionInducedFractionalizationTopological2019}.

For concreteness, let us consider a scheme where, in addition to the $\ket{\mathsf{S}_A}$, $\ket{\mathsf{S}_B}$, and $\ket{\mathsf{P}_0}$ states considered in the main text, we also address two new states, $\ket{\mathsf{P}_1}$ and $\ket{\mathsf{P}_{-1}}$.
The interaction Hamiltonian is modified in this scenario, with new terms coupling the new $\mathsf{P}$ states, given by
\begin{align}
    H_{\rm int} = \sum_{\eta \in \{ A, B \}} \left[ |\mu_\eta|^2 \ket{\mathsf{S}_\eta \mathsf{P}_0}\bra{\mathsf{P}_0 \mathsf{S}_\eta} - \sum_{q \in \{-1, 1\}} |\mu_{\eta,q}|^2\ket{\mathsf{S}_\eta \mathsf{P}_q}\bra{\mathsf{P}_q \mathsf{S}_\eta} \right] + \mathrm{h.c.},
\end{align}
where $\mu_{\eta,q}$ denotes the transition dipole moment between $\ket{\mathsf{S}_\eta}$ and $\ket{\mathsf{P}_q}$.
We consider the case where we apply driving fields to hybridize $\ket{\mathsf{S}_A}$ and $\ket{\mathsf{P}_1}$, and $\ket{\mathsf{S}_B}$ and $\ket{\mathsf{P}_{-1}}$, yielding new single-atom eigenstates
\begin{alignat}{2}
    \ket{\bar{\mathsf{S}}_A} &= \sqrt{a}\ket{\mathsf{S}_A} &&- e^{i\phi} \sqrt{1-a}\ket{\mathsf{P}_1} \\
    \ket{\bar{\mathsf{P}}_1}&= \sqrt{a}\ket{\mathsf{P}_1} &&+ e^{-i\phi}\sqrt{1-a}\ket{\mathsf{S}_A} \\
    \ket{\bar{\mathsf{S}}_A} &= \sqrt{b}\ket{\mathsf{S}_B} &&- e^{i\phi} \sqrt{1-b}\ket{\mathsf{P}_{-1}} \\
    \ket{\bar{\mathsf{P}}_{-1}}&= \sqrt{b}\ket{\mathsf{P}_{-1}} &&+ e^{-i\phi}\sqrt{1-b}\ket{\mathsf{S}_B}.
\end{alignat}
Now rewriting $H_{\rm int}$ in this basis, we note that all energy-conserving terms conserve the number of $\ket{\bar{\mathsf{P}}_{\pm1}}$ excitations.
Thus, if we prepare our state entirely in the $\{ \ket{\bar{\mathsf{S}}_A}, \ket{\bar{\mathsf{S}}_B}, \ket{\mathsf{P}_0}\}$ manifold, the states $\ket{\bar{\mathsf{P}}_{\pm1}}$ can be neglected.
Taking the secular approximation, we have
\begin{align}
    H_{\rm int} = 
    &\abs{\mu_{A}}^2 \; a \left(\ket{\bar{\mathsf{S}}_A \mathsf{P}_0} \bra{\mathsf{P}_0 \bar{\mathsf{S}}_A} + \mathrm{h.c.}\right)\nonumber \\
    +&\abs{\mu_{B}}^2 \; b \left(\ket{\bar{\mathsf{S}}_B \mathsf{P}_0} \bra{\mathsf{P}_0 \bar{\mathsf{S}}_B} + \mathrm{h.c.}\right)\nonumber \\
    -&2 \abs{\mu_{A1}}^2 \; a(1-a) \left(\ket{\bar{\mathsf{S}}_A \bar{\mathsf{S}}_A} \bra{\bar{\mathsf{S}}_A \bar{\mathsf{S}}_A}\right)\nonumber\\
    -&2 \abs{\mu_{B \; \textrm{-1}}}^2 \; b(1-b) \left(\ket{\bar{\mathsf{S}}_B \bar{\mathsf{S}}_B} \bra{\bar{\mathsf{S}}_B \bar{\mathsf{S}}_B}\right)\nonumber \\
    -&\left[\abs{\mu_{B\;\textrm{-}1}}^2 \; b(1-a) + \abs{\mu_{A1}}^2 \; a(1-b)\right] \left(\ket{\bar{\mathsf{S}}_B \bar{\mathsf{S}}_A} \bra{\bar{\mathsf{S}}_A \bar{\mathsf{S}}_B} + \mathrm{h.c.}\right).
\end{align}
Rewriting this Hamiltonian in the bosonic language, we obtain a hardcore boson model of the form
\begin{align}
H_{\rm tJVW} =  \sum_{i,j} \Big[ t_a (a^\dag_i a_j + h.c.) + t_b (b^\dag_i b_j + h.c.) &+ J_z S^z_i S^z_j + J_{\perp} (S^+_iS^-_j + h.c.) \nonumber \\[-1.5ex]
& + V n^t_i n^t_j + W (n^t_i S^z_j + S^z_i n^t_j ) \Big],
\end{align}
where the operators are $S^z_i = (n^a_i - n^b_i)/2$, $S^+_i=a^\dag_i b_i$, $n^t_i = (n^a_i + n^b_i)$, and the couplings are
\begin{align}J_z/4 = V  &=-\frac{1}{2}\left( \abs{\mu_{A1}}^2  a(1-a) + \abs{\mu_{B \; \textrm{-1}}}^2  b(1-b) \right),\\ W&=-\left( \abs{\mu_{A1}}^2  a(1-a) - \abs{\mu_{B \; \textrm{-1}}}^2  b(1-b) \right),\\ t_A&= \abs{\mu_{A}}^2 a; \quad t_B= \abs{\mu_{B}}^2 b,\\
J_\perp &= -\left(\abs{\mu_{B\;\textrm{-}1}}^2 \; b(1-a) + \abs{\mu_{A1}}^2 \; a(1-b)\right).
\end{align}
Moreover, if a F\"orster resonance occurs, we add an additional coupling to the interaction Hamiltonian,
\begin{align}
    H^{}_{\mathrm{{\rm U}(1)}} = H^{}_{\rm int} &+ \bigg[\mu^{}_{A} \mu^{}_{B}(\ket{\mathsf{S}_B \mathsf{S}^{}_A}\bra{\mathsf{P}^{}_0 \mathsf{P}^{}_0} + \ket{\mathsf{S}^{}_A \mathsf{S}^{}_B}\bra{\mathsf{P}^{}_0 \mathsf{P}^{}_0}) \nonumber \\
    & -\sum_{q \in \{\textrm{-}1, 1\}} \mu^{}_{B\;\textrm{-}q} \mu^{}_{Aq}\ket{\mathsf{S}^{}_A \mathsf{S}^{}_B}\bra{\mathsf{P}^{}_q \mathsf{P}^{}_{\textrm{-}q}}
    + \mu^{}_{Bq} \mu^{}_{A\;\textrm{-}q}\ket{\mathsf{S}^{}_B \mathsf{S}^{}_A}\bra{\mathsf{P}^{}_q \mathsf{P}^{}_{\textrm{-}q}}
    + \mathrm{h.c.} \bigg].
\end{align}

\bibliography{refs,zotero_refs}